\documentclass[trackchanges]{aastex62}

\received{--}
\revised{--}
\accepted{--}
\submitjournal{PASP}

\graphicspath{{./}{figures/}}

\begin{document}

\title{ROME/REA: A gravitational microlensing search for exo-planets beyond the snow-line on a global network of robotic telescopes}

\correspondingauthor{Yiannis Tsapras}
\email{ytsapras@ari.uni-heidelberg.de}

\author[0000-0001-8411-351X]{Yiannis Tsapras}
\affiliation{Zentrum f{\"u}r Astronomie der Universit{\"a}t Heidelberg, 
Astronomisches Rechen-Institut,
M{\"o}nchhofstr. 12-14,
69120 Heidelberg, Germany}

\author{R.A. Street}
\affiliation{Las Cumbres Observatory Global Telescope Network,
6740 Cortona Drive, suite 102, Goleta,
CA 93117, USA}

\author{M. Hundertmark}
\affiliation{Zentrum f{\"u}r Astronomie der Universit{\"a}t Heidelberg,
Astronomisches Rechen-Institut,
M{\"o}nchhofstr. 12-14,
69120 Heidelberg, Germany}

\author{E. Bachelet}
\affiliation{Las Cumbres Observatory Global Telescope Network, 
6740 Cortona Drive, suite 102, Goleta,
CA 93117, USA}

\author{M. Dominik}
\affiliation{{SUPA, School of Physics \& Astronomy,
University of St Andrews, North Haugh,
St Andrews KY16 9SS, UK}}

\author{V. Bozza}
\affiliation{Dipartimento di Fisica "E.R. Caianiello", Universit{\`a} di Salerno, Via Giovanni Paolo II 132, 84084 Fisciano, Italy.}
\affiliation{Istituto Nazionale di Fisica Nucleare, Sezione di Napoli, Via Cintia, 80126, Napoli, Italy.}

\author{A. Cassan}
\affiliation{Institut d'Astrophysique de Paris, Sorbonne Universit\'e, CNRS, UMR 7095, 98 bis boulevard Arago, 75014 Paris, France}

\author{J. Wambsganss}
\affiliation{Zentrum f{\"u}r Astronomie der Universit{\"a}t Heidelberg,
Astronomisches Rechen-Institut,
M{\"o}nchhofstr. 12-14,
69120 Heidelberg, Germany}
\affiliation{International Space Science Institute (ISSI), Hallerstrasse 6, 3012 Bern, Switzerland}

\author{K. Horne}
\affiliation{{SUPA, School of Physics \& Astronomy,
University of St Andrews, North Haugh,
St Andrews KY16 9SS, UK}}

\author{S. Mao}
\affiliation{Department of Astronomy and Tsinghua Centre for Astrophysics, Tsinghua University, Beijing 100084, China}
\affiliation{National Astronomical Observatories,
Chinese Academy of Sciences,
100012 Beijing, China}

\author{W. Zang}
\affiliation{Department of Astronomy and Tsinghua Centre for Astrophysics, Tsinghua University, Beijing 100084, China}

\author{D.M. Bramich}
\affiliation{New York University Abu Dhabi, PO Box 129188,
Saadiyat Island, Abu Dhabi, UAE}

\author{A. Saha}
\affiliation{National Optical Astronomy Observatory, 950 North Cherry Ave.,
Tucson, AZ 85719, USA}


\begin{abstract}
Planet population synthesis models predict an abundance of planets with semi-major axes between 1-10 au, yet they lie at the edge of the detection limits of most planet finding techniques. Discovering these planets and studying their distribution is critical to understanding the physical processes that drive planet formation. 

ROME/REA is a gravitational microlensing project whose main science driver is to discover exoplanets in the cold outer regions of planetary systems. To achieve this, it uses a novel approach combining a multi-band survey with reactive follow-up observations, exploiting the unique capabilities of the Las Cumbres Observatory (LCO) global network of robotic telescopes combined with a Target and Observation Manager (TOM) system. We present the main science objectives and a technical overview of the project, including initial results.
\end{abstract}


\section{Introduction}\label{intro}
The prolific discoveries of planets orbiting distant stars over the past two decades have radically transformed our understanding of the properties of planetary systems \citep{2013Sci...340..572H}. Despite these discoveries, fundamental questions about the formation, physical properties and distribution of exoplanets still remain open due to the dearth of low-mass planets ($M_p < 100 M_{\oplus}$) detected at large orbital distances from their host stars \citep{2003A&A...407..369U,2014ApJ...795...64F}.

Current formation theories postulate that proto-planetary cores form in metal-rich accretion disks surrounding the host star. These proto-planets co-evolve with the disk and can undergo orbital decay due to torque asymmetries in the surrounding disk material \citep{2011ARA&A..49..195A,2012A&A...547A.111M,2013AN....334..589D}. 

The so-called {\it snow-line} is defined as the distance from a star beyond which the disk temperature drops below $\sim$160K and water turns to ice \citep{2011Icar..212..416M}. Theory predicts that beyond the snow-line, the formation of ice grains allows planetary embryos to develop sufficiently massive solid cores and gradually grow by accreting material from the surrounding gaseous disk, transforming them into gas giants. 

Population synthesis simulations predict an abundance of low and intermediate mass planets ($M_p < 100 M_{\oplus}$) beyond the snow-line \citep{2013ApJ...775...42I,2009A&A...501.1139M}, but they remain exceedingly hard to detect and little is known about their properties. Indeed, even all of the transiting planets discovered by the highly accomplished Kepler space mission are far too close to their host stars to begin investigating these predictions \citep{2013ApJS..204...24B,2010Sci...327..977B}. 

A recent statistical analysis of microlensing planets by \citet{2018ApJ...869L..34S} found a discrepancy between the microlensing results and the number of intermediate-mass giant planets predicted by planet population synthesis simulations. Since the latter often rely on runaway gas accretion to produce gas giants, a standard assumption in core accretion theory, the microlensing results imply that there may be physical processes involved in giant planet formation that have been overlooked or underestimated in existing models. The significance of this apparent discrepancy between theory and observations has wide-ranging implications and can only be assessed by concentrated efforts to increase the sample of exoplanets discovered beyond the snow-line.

Besides the tantalising possibility of discovering cold Earths, finding these planets is therefore crucial in understanding the physical processes that drive planet formation \citep{2012ARA&A..50..411G}. Each planet detection method is sensitive to a different domain of the planet distribution in mass and distance and the emerging pattern provides a basis for testing and developing our understanding of how planets form and how their orbits evolve. 

{\it Gravitational microlensing} detects planets by measuring how light rays from a background source star bend as they pass through the gravitational field of an intervening planetary system (lens) on their way to our telescopes \citep{1991ApJ...374L..37M,1992ApJ...396..104G}. What is actually observed during a microlensing event is a gradual increase in the brightness of the source star as the lens appears to move closer to it on the plane of the sky, followed by a gradual dimming back to its normal brightness as the lens moves away.The gravitational influence of planets in orbits of a few au around the lens star, typically an M or K-dwarf, can further bend the light rays coming from the source star. The presence of these unseen planets is then revealed through the detection of brief but intense changes in the measured brightness. This method opens up a unique window to the population of low-mass exoplanets at or beyond the snow-line, which is unavailable to other detection methods \citep{2016MNRAS.457.1320T,2012Natur.481..167C,2010ApJ...720.1073G}. 

Microlensing is thus the fastest, cheapest\footnote{Ground-based observations every 15 minutes using modest 1m-class telescope facilities are sufficient to detect the signals of planets as small as the Earth \citep{2006A&G....47c..25D}.} and most time-efficient way to probe the population of exoplanets at moderate to large separations (1-10 au), exploring the region where rock/ice cores are predicted to grow and undergo runaway gas accretion\footnote{Provided the core mass has grown enough to be comparable to the mass of the surrounding gaseous envelope.} and providing essential data to back-calibrate planet population synthesis models. Furthermore, the method is uniquely sensitive to planetary systems at distances of several kilo-parsecs, affording us a view of the true Galactic population of planets \citep{2018Geosc...8..365T,2018arXiv181203137S}.

In Section \ref{project}, we discuss the motivation for the project and how it compares with and can benefit existing efforts within the microlensing community. A technical overview of the facilities and instruments used for the project is given in Section \ref{technical}. The observing strategy is described in Section \ref{observing}. Expected yields based on a simulation of a full observing season are given in Section \ref{simulation}. We conclude with the presentation of some initial results in Section \ref{preliminary}.

\section{PROJECT MOTIVATION}\label{project}
ROME/REA\footnote{Robotic Observations of Microlensing Events/Reactive Event Assessment} is an observational science project running on the global robotic telescope network of the Las Cumbres Observatory (LCO) with the aim of discovering exoplanets beyond the snow-line of their host stars using the technique of gravitational microlensing.

Microlensing event rates are highest in a $\sim$4 square degree area close to the Galactic centre due to the sheer number of available source and lens stars \citep{2013ApJ...778..150S}. Although this area is in the observing footprint of existing surveys, their observations are typically obtained in a single band and only occasionally (weekly or monthly) in a second band, making it difficult to characterise the source star \footnote{Source stars are generally too faint for spectroscopy.} \citep{2015AcA....65....1U,2016JKAS...49...37K,2001MNRAS.327..868B}.

Source star characterisation plays an important part in interpreting a microlensing event since the angular size of the source can be used to estimate the angular Einstein radius of the lens and thus its mass. The relevant equation is  $\theta_S = \rho \theta_E$, where $\theta_S$ is the angular source size, $\theta_E$ the angular Einstein radius and $\rho$ the angular size of the source normalised to the angular Einstein radius of the lens \citep{1994ApJ...430..505W,2004ApJ...603..139Y}. The latter is obtained during the model fitting process when finite source effects are detected in the event light curve. The mass of the lens can then be derived from
\begin{equation}
    M_L = \frac{c^2 \theta_S^2}{4G\rho^2\left(\frac{1}{D_L}-\frac{1}{D_S}\right)},
\end{equation}
where $D_{L}$ is the distance to the lens and $D_S$ the distance to the source.
Therefore, knowledge of the distance and spectral type of the source (and hence its radius) is essential in order to constrain the mass of the lens, and if the distance to the lens can be inferred from the relative lens-source parallax, $\pi_{rel}$ \citep{1992ApJ...392..442G}, then the lens mass can be uniquely determined. For the faint (I$\sim$15-19mag) stars that populate our target fields, multi-band photometry provides this crucial information about the source. Furthermore, since the microlensing effect is achromatic, regular observations in multiple bands can also be used to distinguish between light coming from the source star, which contains the microlensing signal, and blended light from faint stars at roughly the same position on the sky as the source, which dilutes that signal by different amounts in different bands. We now turn to how this is done in practice.

The ROME/REA project is conducted exclusively on the 1m telescopes that belong to LCO's network in the southern hemisphere. We use a novel observing strategy that relies on LCO's robotic telescope observing framework and which complements the scientific goals of other microlensing surveys by ensuring that the source stars within our survey footprint are well characterised and therefore that the physical parameters of the lenses can be well determined. We achieve this by extracting the source type through the use of time series data in three observing bands, which also allows us to constrain how blended the event is. 

Three bands is the minimum required to enable blend analysis using a colour-colour diagram. On this diagram, an unblended source will appear as a single point for the full duration of the microlensing event, whereas a blended source will trace a curved path as the event evolves\footnote{If a flux ratio diagram (FRD) is used instead, this path will be straight.} because the flux ratio between pure-source and pure-blend fluxes is going to change as it is only the source that experiences magnification due to microlensing. Once the microlensing event has run its course, a line may be fit to the path. If the line is short, then it is an event with low blending. Conversely, if the line is long, the event is heavily blended. The length of the line is therefore a direct indicator of the degree of blending. Furthermore, the colour of the blend itself will change the slope of the line in the colour-colour diagram. 

Information about the stellar type of the source can be obtained by over-plotting the colour differences for (unblended) stars of known spectral types on this diagram and identifying their (fixed) locations. The location of the source, after accounting for extinction, can then be directly compared with these known positions in the diagram and the source star can be associated with a particular spectral type. The angular radius of the source $\theta_S$ can then be estimated using the relationships derived in \citet{2014AJ....147...47B}.

A detailed first demonstration of the strategy and associated results can be found in \citet{2019AJ....157..215S}. In this analysis, the microlensing event investigated (OGLE-2018-BLG-0022) was found to be due to a binary star with individual components of 0.375 $\pm$ 0.020 $M_{\odot}$ and 0.098 $\pm$ 0.005 $M_{\odot}$. For comparison, an independent analysis of the same event by \citet{2019ApJ...876...81H}, relying on the more common approach using only two bands and a more densely sampled light curve, determined the masses of the components to be 0.40 $\pm$ 0.05 $M_{\odot}$ and 0.13 $\pm$0.01 $M_{\odot}$, respectively. This suggests that the three-band approach can potentially improve the accuracy of the mass estimates by at least a factor of two.

Microlensing observations in the past have generally not been associated with extensive log-keeping and metadata structures\footnote{These can contain useful information about the regions of sensitivity of the instruments used, information about local observing conditions at the time of the observation, etc.}, so it is usually not possible to appreciate the reasoning behind human-driven observing decisions or to trace how different instrumental and environmental factors influence the quality of the data \citep{2015ApJ...812..136B}. In addition, surveys have redefined their observing fields and cadences several times over the years, which affects their relative sensitivity to planets at different regions of the sky and can lead to biased estimates of the planet mass function if not carefully accounted for. In our ROME survey, the observing strategy is software-driven and follows a predetermined pattern, while the REA target selection process is also entirely automated (see Section~\ref{observing}). All observing decisions are logged and tracked, allowing us to reconstruct the decision tree that led to any particular observation. Any additional observing requests by members of our team are tagged as such and analysed separately when estimating the planet detection sensitivity of the project as a whole. Thus, knowledge of the conditions under which each observation was performed helps us to better quantify our biases.

The observing sites used to conduct this project are listed in Table~\ref{tab:lco_network}. Apart from the TAC\footnote{TAC: Time Allocation Committee.} allocated observing time, the ROME/REA project is enabled by contributions to the total time budget by the University of St Andrews, Heidelberg University and the Chinese National Academy of Sciences.

\begin{deluxetable*}{lllcl}
\tablecaption{LCO sites and telescopes used by the ROME/REA project\label{tab:lco_network}}
\tablewidth{0pt}
\tablehead{
\colhead{Observing site} & \multicolumn2c{Coordinates} & \colhead{Elevation (m)} &
\colhead{Time zone}
}
\startdata
Siding Spring Observatory & $31^{\circ}16^{'}23.88^{''}$S & $149^{\circ}4^{'}15.6^{''}$E & 1116 & UTC+10 \\
South African Astronomical Observatory & $32^{\circ}22^{'}48^{''}$S & $20^{\circ}48^{'}36^{''}$E & 1460 & UTC+2\\
Cerro Tololo Interamerican Observatory & $30^{\circ}10^{'}2.64^{''}$S & $70^{\circ}48^{'}17.28^{''}$W & 2198 & UTC-3 \\
\enddata
\end{deluxetable*}

\section{Facilities overview}\label{technical}
The telescope network description that follows is an update on the information presented in \citet{2013PASP..125.1031B}. We present here sufficient information to place the following sections in context but refer the interested reader to the original paper for a more detailed description.

LCO is an organisation dedicated to time-domain astronomy. To facilitate this, LCO operates a homogeneous network of 2m, 1m and 0.4m telescopes on multiple sites around the world\footnote{https://lco.global/}, covering both hemispheres. Each observing site hosts between one to five telescopes, which are outfitted for imaging and spectroscopy. The instruments and filters used are the same for all telescopes, allowing `network redundancy', so that observations can be shifted to alternate sites at any time in the case of technical problems or poor weather. 

The 1m telescopes are currently equipped with custom made 4K$\times$4K 15-micron Fairchild CCDs (Sinistro). The pixel scale is 0.389 arcseconds in 1$\times$1 binning mode, giving a field of view of 26.5$\times$26.5 arcminutes. The standard filter loadout is a complete Johnson-Cousins/Bessell set (UBVRI) and a SDSS/PanSTARRS set ($u^\prime g^\prime r^\prime i^\prime z_s^\prime Yw$). We note that ROME observations are performed only in three bands: SDSS-$g^\prime$, SDSS-$r^\prime$ and SDSS-$i^\prime$, while REA observations are only done in SDSS-$i^\prime$. The specified horizon limit of the telescopes is 15 degrees (3.7 airmasses) and their slewing speed is $\sim$6 degrees per second. 

The robotic system is responsible for all telescope functions, including slewing, tracking, auto-guiding, but also controls the functions of the instruments and filter wheels. It features built-in recovery mechanisms to address problems in an automated fashion and has a weather station providing it with continuous information about local conditions so it can shut the enclosure should the humidity or cloud cover exceed the limiting parameter values.

The telescopes are controlled by a single robotic scheduler program, capable of orchestrating complex and highly responsive\footnote{Can respond to and observe any newly alerted target within minutes.} observing programs using the entire network to provide round-the-clock observations of any astronomical target of interest, weather permitting. Sequences of complex observing requests can be submitted programmatically to the network through an application program interface (API). This is flexible enough to allow combinations of different observing strategies within the same science program. For example, a user can write software to specify regular survey observations with a fixed instrumental setup and cadence on the network but also specify conditions for a rapid response function with dynamically adjustable observing times based on the current brightness of the target that will trigger when specific conditions are met.

All observation requests to the network are transferred to a central database, classified by the dynamic scheduler software \citep{2014SPIE.9149E..0ES,2015arXiv150307170L} and handed out to the telescopes in order of scientific importance (assigned by the time-allocation committee to all science programs), observability and internal relative project priority.

\section{Observing strategy}\label{observing}
\subsection{Project setup}
\begin{figure}[ht!]
        \plottwo{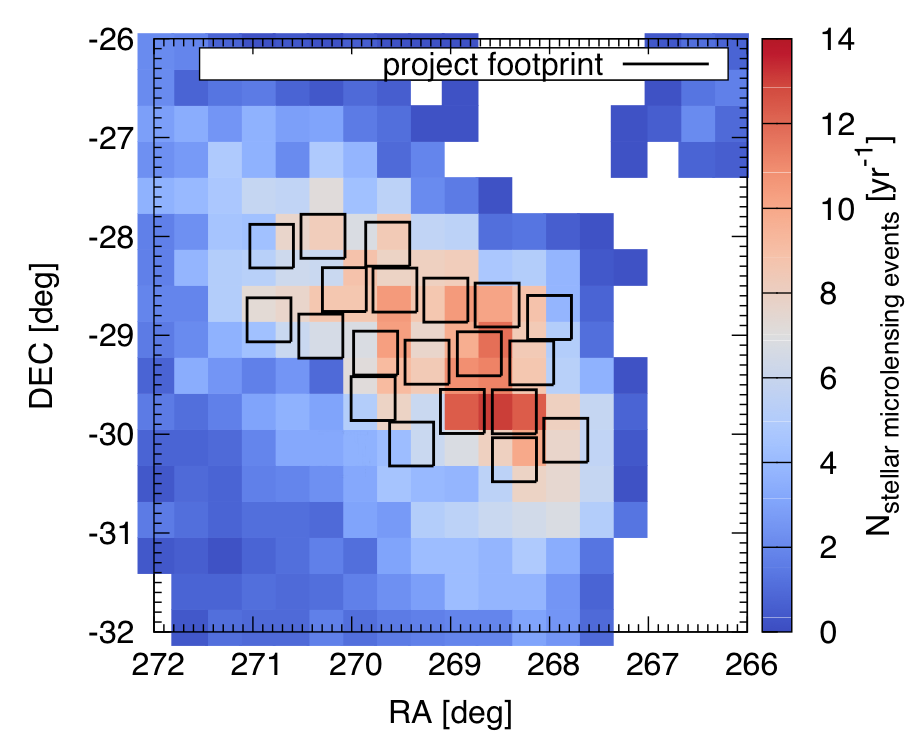}{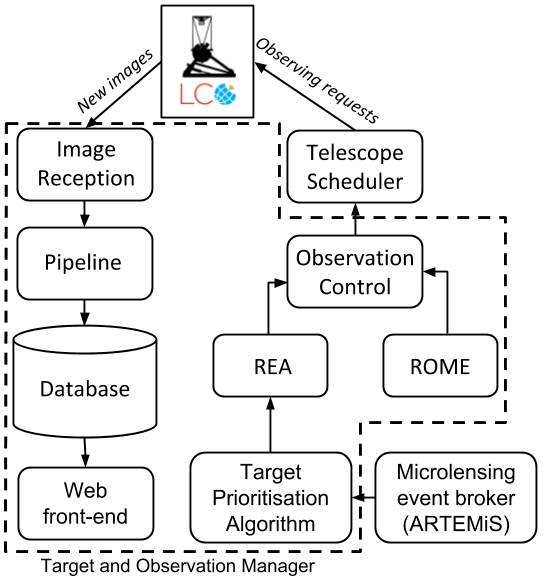}
	    \caption{[Left]:ROME/REA targets twenty crowded stellar fields rich in microlensing events close to the Galactic centre. The exact position of each field on the sky was optimised using an algorithm that avoided regions with very bright stars or very high extinction \citep{2014MNRAS.440.2036D,2013ApJ...769...88N}, while maximising the microlensing event rate and avoiding field overlap. The colours represent the expected number of microlensing events per year. 
	    [Right]: Project system architecture. Both the ROME survey and REA responsive mode automatically submit observing requests to the LCO telescopes through the Telescope Scheduler. New images obtained at the telescopes are promptly identified and processed by our software, and the resulting products are stored in a database. Photometric analysis can be performed either automatically or refined manually.}
	    \label{fig:fields}
\end{figure}
The possibility of having autonomous software agents in control of a science program make the LCO network an indispensable tool for our project. The ROME/REA project combines survey and follow-up observations to maximise the planet detection rate. Microlensing observations take place when the Galactic bulge is observable for extended periods of time from the LCO sites, roughly from 1$^{\mathrm{st}}$ April to 15$^{\mathrm{th}}$ of October each year. Our observing strategy for the ROME survey involves monitoring an event-rich 3.76 square degree area of the sky close to the Galactic centre every seven hours, seamlessly switching the observations between the three southern LCO sites.

Twenty target fields with the highest star counts were selected to maximise the microlensing event rate (see Figure~\ref{fig:fields}), based on the coordinates of OGLE microlensing alerts announced between 2013 and 2015. The field centres are given in Table~\ref{tab:rome_rea_fields}. Regions with very bright stars (Vmag$<$7) that would cause the detector chip to saturate were avoided, as were regions of very high extinction. A sequence of single exposures in each of three bands (SDSS-$g^\prime$, SDSS-$r^\prime$ and SDSS-$i^\prime$) is obtained during every field visit. Although LCO offers a range of filters, these bands were selected because they give the optimal balance between throughput/CCD sensitivity and spectral type classification efficiency. The exposure times are fixed to 300 seconds to reach $\sim$19th magnitude in the SDSS-$i^\prime$ band at an estimated S/N$\sim$50. 
\begin{deluxetable*}{llll}
\tablecaption{ROME/REA field centers\label{tab:rome_rea_fields}}
\tablewidth{0pt}
\tablehead{
\colhead{Field Identifier} & \colhead{RA} & \colhead{Dec}
}
\startdata
ROME-FIELD-01 & 17:51:20.62 & -30:03:38.94\\
ROME-FIELD-02 & 17:58:32.82 & -27:58:41.76\\
ROME-FIELD-03 & 17:52:00.01 & -28:49:10.41\\
ROME-FIELD-04 & 17:52:43.24 & -29:16:42.65\\
ROME-FIELD-05 & 17:53:25.04 & -30:15:28.21\\
ROME-FIELD-06 & 17:53:25.47 & -29:46:22.74\\
ROME-FIELD-07 & 17:54:07.10 & -28:41:37.35\\
ROME-FIELD-08 & 17:54:50.34 & -29:11:12.21\\
ROME-FIELD-09 & 17:55:31.47 & -29:46:13.68\\
ROME-FIELD-10 & 17:56:11.64 & -28:38:38.64\\
ROME-FIELD-11 & 17:56:57.32 & -29:16:18.01\\
ROME-FIELD-12 & 17:57:34.75 & -30:05:57.25\\
ROME-FIELD-13 & 17:58:15.29 & -28:26:32.04\\
ROME-FIELD-14 & 17:59:02.12 & -29:10:46.57\\
ROME-FIELD-15 & 17:59:08.06 & -29:38:21.86\\
ROME-FIELD-16 & 18:00:18.00 & -28:32:15.21\\
ROME-FIELD-17 & 18:03:14.40 & -28:05:52.20\\
ROME-FIELD-18 & 18:01:09.81 & -27:59:54.97\\
ROME-FIELD-19 & 18:01:15.06 & -29:00:30.33\\
ROME-FIELD-20 & 18:03:20.82 & -28:50:35.37\\
\enddata
\tablecomments{Each pointing covers a field of view of 26.5$\times$26.5 arcminutes.}
\end{deluxetable*}

Follow-up observations form part of our reactive REA strategy, which complements our ROME survey and aims to increase the sensitivity to planets below Neptune-mass. At any given time during the observing season, there are about 40-50 concurrent microlensing events. Our algorithm selects those where the probability of detection of planets is highest per time spent observing \citep{2018A&A...609A..55H,2009MNRAS.396.2087H}, typically between two to four events. Extra observing requests for each of these events are then automatically submitted to the robotic observing queue of a single telescope at each southern LCO site with a requested cadence of 1 hour. The software will attempt to avoid queuing reactive REA observations to the telescope performing the ROME survey observations, but that is not always possible or practical. In the event that any of these high-interest targets show anomalous features, the observing cadence can be further reduced to 15 minutes. REA observations are performed in a single band (SDSS-$i^\prime$), with the exposure time evaluated in real-time based on the current brightness of the microlensing event \citep{2008AN....329..248D}, and the pointing is matched to the coordinates of the ROME field the target event appears in. We note in passing that in case the system is unable to perform REA observations, the reasons are usually technical and are independent of knowledge about the events themselves, therefore statistical results would not be biased.
\begin{figure*}[ht!]
    \begin{center}
        \plotone{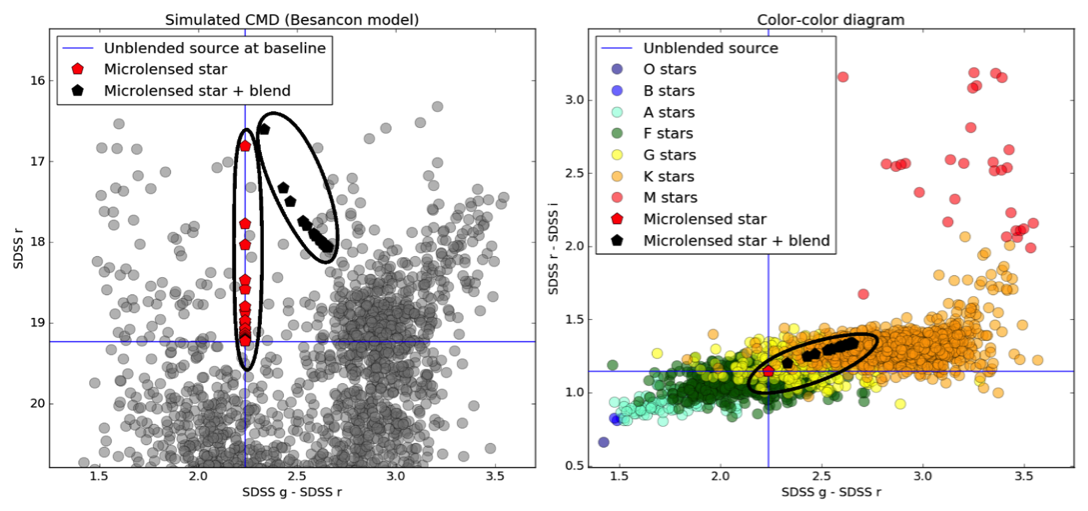}
	    \caption{[Left]A colour-magnitude diagram for a single ROME field, simulated with the Besan\c con model \citep{2003A&A...409..523R}. Red polygons mark the brightness changes of an unblended source star as it gets magnified by a passing foreground lens. Black polygons mark the brightness and colour changes of a blended source star during the course of its magnification. The skewness of this line reveals the degree of blended light contributing to the event. [Right] An unblended microlensed source star would not `move' in this colour-colour diagram since all colours are equally magnified. Source stars with high degrees of blending can be traced through the diagram and reveal the actual source type. Noise has been removed from the microlensing event for illustrative purposes.}
	    \label{fig:colours}
   \end{center}
\end{figure*}

This automated procedure reassesses ongoing events every quarter of an hour and ensures that high-cadence complementary reactive observations are obtained for those microlensing events that show the highest sensitivity to planets, but only if these events happen to lie within the survey areas regularly monitored by ROME. Microlensing events outside our survey footprint do not enter the observing queue, although exceptions can be made in case of particularly remarkable events \citep{2018MNRAS.476.2962N,2019ApJ...871...70D}.

For any ROME or REA observations to be scheduled, the angular distance between the moon and the target field must be greater than 15 degrees. Furthermore, no ROME observations are scheduled during full moon (2 nights/month). A simple overview of the channels of communication between the different parts of the project is shown in the right panel of Figure~\ref{fig:fields}.

Although LCO offers a Target of Opportunity (ToO) rapid response observing option, we do not use it for our project since the scheduler picks up our observing requests within minutes so that a ToO response has been deemed unnecessary. Since REA observations are submitted with higher priority, they are preferentially selected by the scheduler.

We note that the ROME/REA system comprehensively logs and time-stamps all steps that lead to software-driven observing decisions. We are thus able to reconstruct the configuration of the system at any given time and understand why particular observing decisions were made.

\subsection{Target and Observation Manager}
The aforementioned features of the project are embedded in a specialised software framework that is generically referred as a ``Target and Observation Manager'' (TOM) system \citep{2018SPIE10707E..11S}. TOM systems function as proxy astronomers, making observing decisions based on available information. They are typically used to harvest alert streams coming from different surveys, assess the relative importance of these alerts for a given science program, schedule or recommend observations, trigger reduction pipelines, update databases, as well as serve the information back to users in easily accessible formats, such as web feeds. They offer a powerful way to visualise and interact with data through a web browser or a graphical user interface (GUI). TOM systems have been rapidly gaining popularity in the advent of next-generation wide-field surveys, such as ZTF\footnote{Zwicky Transient Facility \citep{2019PASP..131a8002B}} and LSST\footnote{Large Synoptic Survey Telescope \citep{2009arXiv0912.0201L}}, which are expected to produce more than a million alerts of astronomical transients per night. It is humanly impossible to parse such a high volume of alerts on any reasonable timescale and decide the best way to allocate limited observing resources in order to maximise scientific returns. However, a well-designed TOM system can handle all practical aspects. 

The ROME/REA TOM system operates in two modes. The regular monitoring of fixed ROME survey fields is configured and scheduled independently from the REA mode. ROME observations are scheduled daily with a fixed order, cadence and exposure time. For REA, the TOM automatically harvests alerts from a number of surveys, including OGLE, MOA and ZTF. It uses the ARTEMiS\footnote{http://www.artemis-uk.org/} microlensing alert broker \citep{2007MNRAS.380..792D,2008AN....329..248D}, designed to identify and track ongoing anomalous microlensing events, passing all relevant information to an automated target selection and prioritisation algorithm \citep{2018A&A...609A..55H}. For the targets selected, the algorithm evaluates the exposure times to be used based on the predicted current brightness of the event and submits groups of observing requests to the LCO telescopes without requiring manual intervention. 

Our TOM system continuously runs another piece of software in the background which identifies recently acquired images for the ROME/REA project in the LCO image archives and automatically transfers them back for processing. Incoming images automatically trigger a customised photometric pipeline to produce or update light curves in the ROME/REA database. 
A web front-end is also provided as part of the TOM for team members to track and assess the performance of the algorithms at any given time, to identify potential problems at a glance, and to visualise the light curve of any given target of interest.

\section{Simulating a season}\label{simulation}
In order to estimate the expected yield of our project, we simulated a full microlensing observing season as it would be observed with the strategy outlined in section~\ref{observing}. To generate the sample of microlensing events, we used the published parameters of microlensing events detected during the 2015 microlensing season by the OGLE-IV survey \citep{2015AcA....65....1U}, provided their coordinates matched our ROME survey footprint. The light curves of these events were then sampled based on our observing strategy. The simulation was done using the open-source pyLIMA microlensing modeling software \citep{2017AJ....154..203B}. It included losses due to weather, based on historical weather data at the sites of the LCO telescopes, and observing limits set by the proximity of the target field to the moon. The noise model was calibrated using data from previous microlensing observing seasons at LCO with a similar technical setup, specifically the 2016 season of the RoboNet microlensing project \citep{2009AN....330....4T}. 

For each event, the planet detection probability was evaluated assuming each star has one planet between 0.5-10 au (uniform in log($a$)). The signal of an artificial planet with a mass derived from the \citet{2012Natur.481..167C} planet mass function was injected into the light curve and we evaluated whether it would be detected or missed given our sampling. To consider a planet ``detected'', we required at least seven consecutive observations during the planetary anomaly \citep{2010AN....331..671D}, with each corresponding point deviating by more than 3$\sigma$ from the unperturbed single-lens microlensing light curve. The general procedure is described in \citet{2018A&A...609A..55H}. In effect, we do this in order to avoid mis-estimating crucial statistical properties due to small samples. For the simulation, all high magnification events were already selected and being ``observed'' in REA 1-hour mode. ROME ``observations'', one in each band every seven hours, were also included in this assessment. We did not include a REA 15-minute observing cadence in the simulations because the automatic alert assessment system produced many false positives, which, since these observations are very time-costly, would deplete our allocated time very quickly.

Our simulation results can be summarised as follows: Assuming that the technical performance of the telescope network remains stable, we expect our project to yield at least $\sim$10 new cool planet discoveries in its three-year lifetime, if such planets are as abundant as theory predicts. We estimate that $\sim$79\% of the planets we find will have masses between 30 M$_\oplus$ and 10 M$_{\mathrm{Jup}}$, and $\sim$21\% between 5 and 30 M$_\oplus$. The exact number and type of planets will depend on the true underlying planet population statistics and our detection efficiency, but any findings in this planetary parameter space are of great scientific interest since recent statistical results from microlensing \citep{2016ApJ...833..145S,2018ApJ...869L..34S} identify a break in the mass-ratio function with few planets detected in the low-mass range ($M_p \le 20 M_{\oplus}$). If confirmed, this result would imply that planet formation processes are not as efficient in producing smaller mass planets as current planet population synthesis models suggest \citep{2013ApJ...775...42I}.

Although most of these planets are expected to be identified separately by other surveys, the ability to characterise the source star using observations in three bands is particular to this project.


Figure~\ref{fig:colours} was generated from our simulation using the Besan\c con model of stellar population synthesis of the Galaxy and demonstrates the usefulness of multi-band photometry, as previously described in Section~\ref{project}. We used the underlying distributions to draw the lens and source distances for our simulations. The left panel shows a colour-magnitude diagram (CMD) for a single ROME survey field and how our program can distinguish between different degrees of blending. On the right, the advantage of colour-colour information is illustrated: highly blended source stars can be identified and traced through this diagram as the microlensing event evolves, thereby revealing their stellar type. Our simulations also showed that it is possible to obtain useful data in all three bands for about a third of our total star sample. It is difficult in practice to extract accurate measurements for the fainter events close to our detection limits, especially in SDSS-$g^\prime$, as well as for very short events.

\section{Initial results}\label{preliminary}

\subsection{A new photometric pipeline}
\begin{figure*}
        \plotone{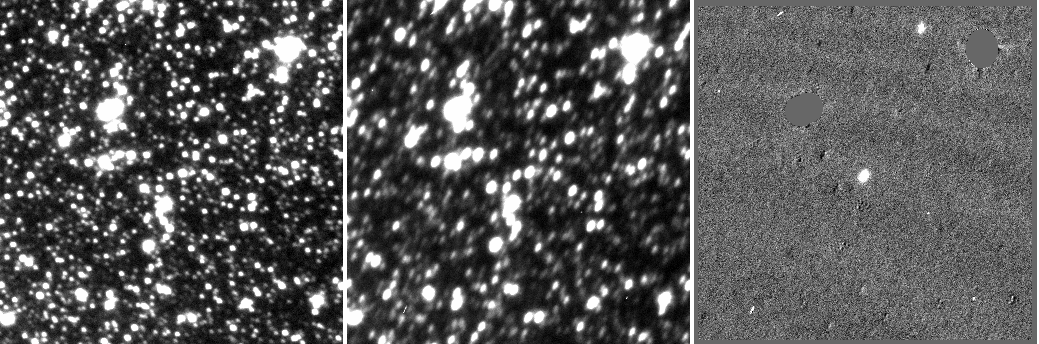}
	    \caption{Image subtraction results. A reference image, shown on the left, obtained under good observing conditions is adjusted to match the orientation and seeing distortions of a target image, shown in the centre, obtained at a different epoch. The two are subtracted and the resulting difference image is presented in the right panel. The image thumbnail is centred on event OGLE-2019-BLG-0171.}
	    \label{fig:diffimg}
\end{figure*}
Image subtraction involves the subtraction of all constant-in-time features from time-series of astronomical images of the same field, using them to self-calibrate the photometric scale \citep{1996AJ....112.2872T,1998ApJ...503..325A}. This is done by constructing a reference image, which can be a single image, or a combination of a number of images, obtained under good observing conditions, and adjusting it to match the observing conditions of every other image taken at different epochs. The adjustment for a single pair of images involves a convolution that registers the images, blurs them to match the atmospheric seeing, and scales them to match the atmospheric transmission \citep{2000A&AS..144..363A}. The adjusted reference image is then subtracted from the target image, removing all constant-in-time features and leaving behind only residual noise and the signals of sources that have varied between the two images. The brightness fluctuations of a variable object, such as a microlensing target, can then be measured on the full set of difference images.

\citet{2008MNRAS.386L..77B} presented an alternative method to standard image subtraction for determining the convolution kernel that matches pairs of images of the same field. The technique involves defining the kernel as a discrete pixel array which, at the cost of some computational speed, can deal sufficiently well with asymmetric point-spread functions and small image misalignments. 

Image subtraction software, developed in IDL\footnote{Interactive Data Language.} and used for the RoboNet microlensing follow-up project \citep{2009AN....330....4T}, employed this technique to produce consistently reliable precision photometry optimised for the extraction of single microlensing event light curves. However, the expected computational speed drop for a project of the scale of ROME/REA necessitated the development of new algorithms, more suited to the particular needs of the project. Specifically, we want each incoming $\sim4$k$\times$4k pixel image to be automatically picked up and processed by our photometric pipeline and new photometric points extracted for all stars in a timely fashion. Each updated light-curve can then be used to search for and assess ongoing microlensing events in real-time using our machine-learning event classifier.

Our new photometric pipeline (pyDANDIA\footnote{https://github.com/pyDANDIA/pyDANDIA}) is written in the Python programming language and is optimised for the cameras of the LCO network. We achieved improved computational speed by trading some photometric accuracy at the fainter end of the magnitude distribution. The results of a detailed evaluation of the performance of the pipeline will be published in a separate paper. A typical example of image subtraction is presented in Figure~\ref{fig:diffimg}.

\subsection{Producing a star catalogue}

A triplet of high-quality template images that have been obtained as part of the same exposure sequence, one in each observing band, are used to generate a star catalogue for each of the twenty ROME survey fields. This is done to ensure that the measured colours are derived from data taken approximately contemporaneously and under as similar observing conditions as possible.
\begin{figure*}
\gridline{\fig{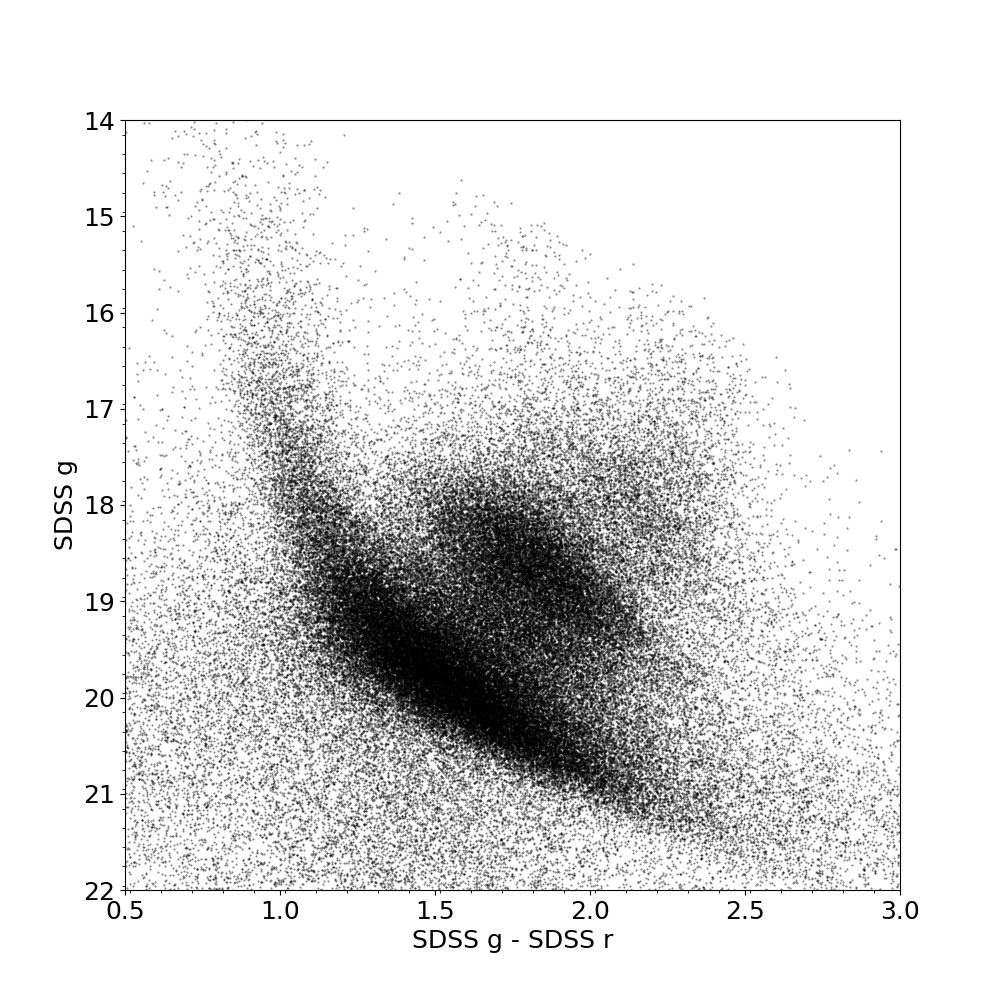}{0.45\textwidth}{(a)}
\fig{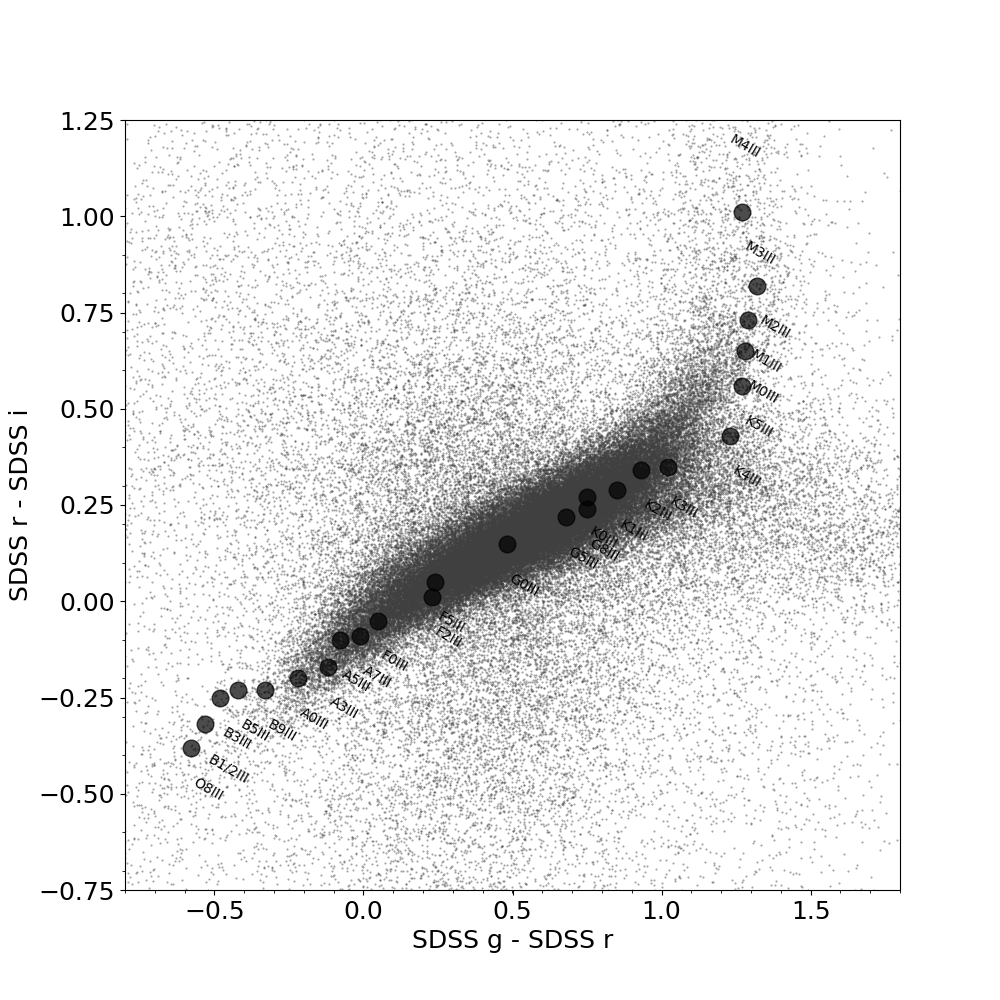}{0.45\textwidth}{(b)}
}
\caption{(a) Color-Magnitude diagram of ROME-FIELD-16. Measurements for 125593 stars were cross-matched between the SDSS-$g^\prime$, SDSS-$r^\prime$ and SDSS-$i^\prime$ bands. (b) Colour-Colour diagram of ROME-FIELD-16. The different spectral types marked by the grey dots are obtained from the atlas of synthetic spectra of \citet{1985ApJS...59...33P}. (Values not corrected for extinction.)}
\label{fig:colors}
\end{figure*}

Individual sources are identified and measured on each of the three template images using a custom implementation of the DAOFIND algorithm \citep{1987PASP...99..191S}, followed by a refinement of the World Coordinate System (WCS) solution using the Gaia DR2 catalogue and photometric calibration by cross-matching with the VPHAS+ Public Survey catalogue\footnote{http://www.vphas.eu/data.shtml}. Sources are then cross-matched between all three images and a final catalogue of stars is produced for the particular ROME field\footnote{We note that inclusion in the final catalogue requires that a source is detected in at least one of the three template images.}.

The calibrated star positions and magnitudes in these catalogues remain fixed as long as the template images used for the photometric reduction do not change. We expect that the catalogues will need to be revised two or three times during the course of the project once stacked template images have been constructed, offering improved sensitivity to fainter magnitudes.

Figure~\ref{fig:colors} shows a colour-magnitude diagram from our initial reduction of a single ROME field. The mean number of stars detected per field, averaged over all ROME fields, is $\sim$150 thousand down to magnitude $I\sim$19. We note that this value is based on our initial analysis performed using a single template image in each band. By generating stacked template images at the end of the project, we expect the magnitude sensitivity to improve by between 0.5 to 1 magnitudes, and the final catalogue release will contain multi-band photometry for about 3 million stars.

When analysing individual microlensing events, our colour analysis only uses stars within 2 arcmin from the event coordinates to generate these diagrams, so as to minimise the effect of differential extinction across the field of view. Correcting for extinction then involves identifying the location of the Red Clump (RC) stars on the colour-magnitude diagram, measuring its offset from known values \citep{2013ApJ...769...88N,2018A&A...609A.116R} and using it to estimate of photometric properties of the source\footnote{Under the assumption that the source is at the same distance and suffers the same extinction as the RC.}. Furthermore, since the magnification of an event can be assumed to be approximately the same in each set of SDSS-$g^\prime$,-$r^\prime$,-$i^\prime$ ROME images, taken within 15 minutes of each other, it is possible to measure the colour of the source independent of the model by calculating the colour-colour slope for different sets of pass-bands and for different magnifications. An end-to-end colour analysis of microlensing event OGLE-2018-BLG-0022 using ROME/REA observations exclusively was recently presented in \citet{2019AJ....157..215S}, where this procedure is described in detail.

\begin{figure*}
\gridline{\fig{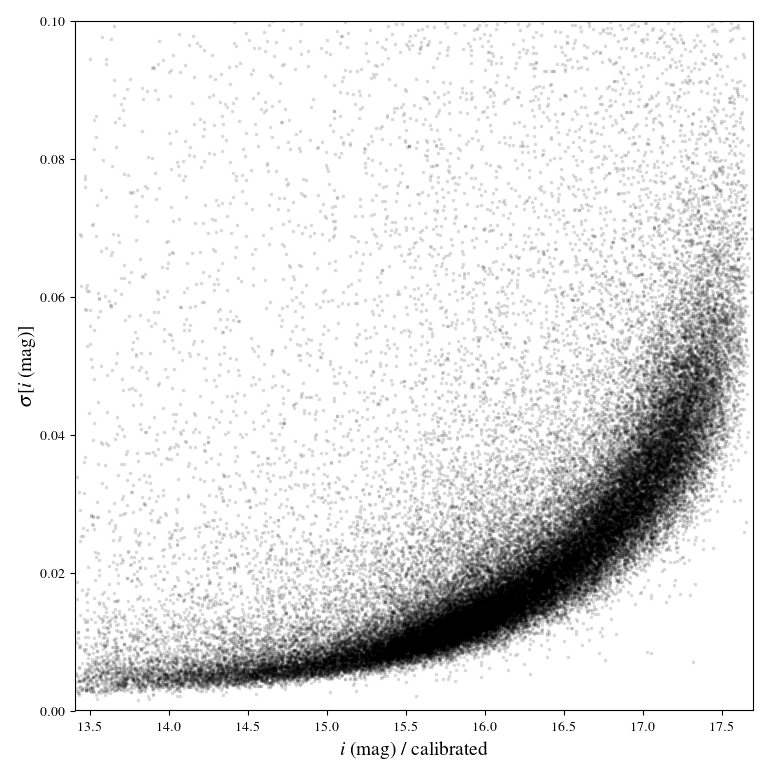}{0.45\textwidth}{(a)}
\fig{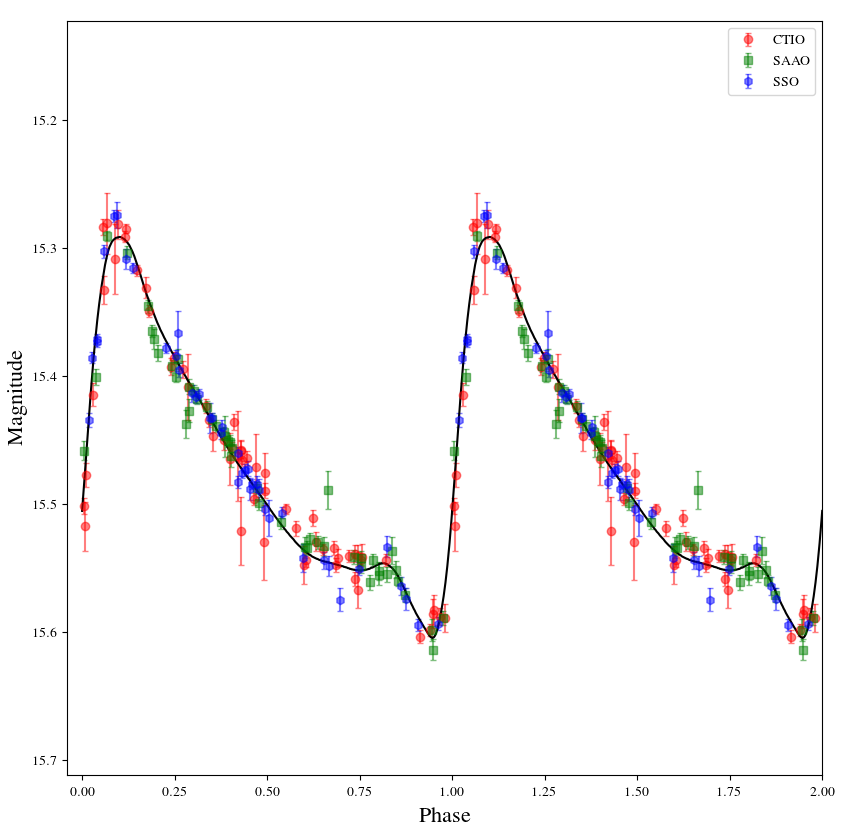}{0.45\textwidth}{(b)}
}
\caption{(a) Accuracy of ROME/REA SDSS-$i^\prime$ photometry from an initial analysis of SAAO observations.  (b) Sample SDSS-$i^\prime$ phase-folded light curve of RR Lyrae variable OGLE-BLG-RRLYR-07770, as observed from LCO telescopes in Chile (CTIO), South Africa (SAAO) and Australia (SAO). The solid black line shows our best RR Lyrae model fit to the data, following the method described in \citet{2017MNRAS.465.2489T}.}
\label{fig:rrlyr}
\end{figure*}

Figure~\ref{fig:rrlyr}(a) presents the photometric accuracy in SDSS-$i^\prime$ for 68293 stars in a 3k$\times$3k sub-region of ROME field 5. Such diagrams are part of a set of diagrams automatically constructed for each ROME field and are used to evaluate the quality of the photometry. 

Our data set contains observations in three bands for thousands of variable stars. For example, Figure~\ref{fig:rrlyr}(b) shows the phase-folded light curve of known RR Lyrae variable star OGLE-BLG-RRLYR-07770 (RA/Dec(J2000): 17:56:01.90, -29:51:45.3, I=15.461, V=17.591). For clarity, only SDSS-$i^\prime$ observations from the three southern LCO sites are displayed.

\subsection{Machine-learning event identification}
Our team has developed an efficient machine-learning classifier that uses the Random Forest algorithm to identify microlensing signals in time-series data\footnote{https://github.com/dgodinez77/LIA}.
The classifier is designed for flexibility of use and has been successfully applied on OGLE-II survey data, recovering $\sim$95\% of known microlensing events. In addition, it has been tested on archival data from the Palomar Transient Factory (PTF) optical wide-field survey and alert-stream data from the ZTF survey.

\begin{figure*}
        \plotone{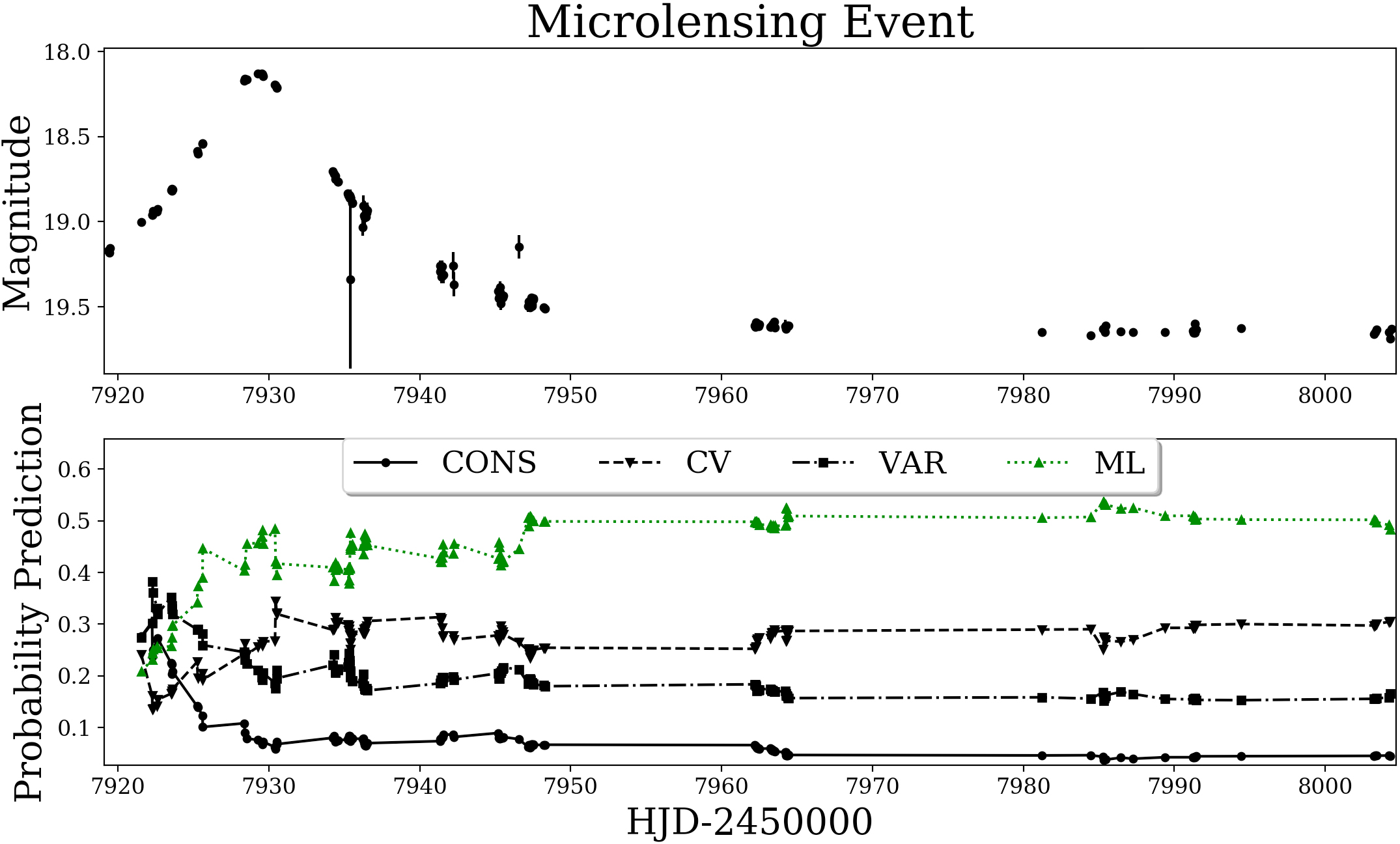}
	    \caption{[Top]: The light curve of a microlensing event identified in a test run of our Random Forest classifier on a subset of ROME data. [Bottom]: Drip-feeding analysis reveals at which epoch the classifier would have identified this as a microlensing event (t=7925). Four different classes were considered: CONS for constant star, CV for cataclysmic variable, VAR for RR Lyrae or Cepheid and ML for microlensing.}
	    \label{fig:mlrun}
\end{figure*}

The top part of Figure~\ref{fig:mlrun} presents the light curve of one of the microlensing events recovered by running the classifier on the initial photometry of a single ROME/REA field. For the purposes of this exercise, light curves containing fewer than 5 photometric measurements were removed from the sample. The bottom panel of the figure demonstrates how the classifier performs in real-time. Four different classes of variability are considered: CONS (constant star), CV (cataclysmic variable), VAR (RR Lyrae or Cepheid) and ML (microlensing). Every incoming data point changes the relative probability of the light curve belonging to each of the classes. The microlensing `detection' happens at the epoch when the green line crosses above the others.

The methodology, features considered and technical details of the classifier can be found in \citet{daniels_paper}.


\section{Summary}\label{summary}
The ROME/REA project uses the Las Cumbres Observatory network of robotic telescopes to detect and characterise the brief signals cold exoplanets can produce in the light curves of microlensing events. 

The ROME survey is the first extensive campaign to map an area of $\sim$4 degrees close to the Galactic centre in multiple bands with a cadence of 7 hours. Observations in three bands are useful for characterising the spectral type of the source stars and identifying contaminating blended light. With this information at hand, it is possible to place stronger constraints on the mass of the lens, thereby improving the precision of the mass estimate of any planetary companions.

The reactive observing REA campaign, with its default 1 hour observing cadence, complements the  ROME survey by targeting specific microlensing events where the probability of detecting the signals of planetary companions to the main lens star are highest. This higher cadence offers enhanced sensitivity to planets with smaller masses (less than 10 $M_{\oplus}$) and, when anomalous features are detected, can be further adjusted to sample the light curve every 15 minutes.

For the great majority of high-magnification events in the ROME/REA sample, there will be sufficiently dense coverage of the light curve to characterise the physical nature of the event, thereby leading to either entirely new discoveries or independent validations of newly discovered events by other microlensing surveys. In addition, even in the cases where ROME/REA observations are too sparse for characterisation, multi-band data will still be of great use to other surveys in constraining the stellar type of the source and a combined analysis of all available data can lead to a better understanding of the properties of the lens \citep{2012ApJ...754...73B,2016AJ....152..125B,2019AJ....157..215S,2019MNRAS.tmp.1368T}.

A public release of the ROME/REA photometric catalogue in three bands for about 3 million stars brighter than $I\sim 20$mag in our observing fields is planned after the end of the project. The final star catalogue will be a valuable side-product of our survey and will be freely available to the astronomical community. All data products and codes developed during the course of this project are to be released under a public license, including our new image subtraction pipeline, and user-friendly software to model microlensing event light curves (pyLIMA\footnote{https://github.com/ebachelet/pyLIMA}\citep{2017AJ....154..203B}, muLAn\footnote{https://github.com/muLAn-project/}\citep{muLAn}). Studies of variable stars, Galactic extinction, transiting exoplanets and stellar remnants are some potential secondary science by-products of our final data release.

The dynamic observing techniques we have developed within the framework of the LCO network rely on parsing incoming alert streams and modifying observing proposals in real-time. They introduce new targets, adjust relative priorities between targets, alter the observing cadence and exposure times, always using the latest available information. Such intelligent robotic observing agents are of great interest to the wider transient astronomical community as new methods are urgently sought to handle the high-volume alert streams from upcoming wide-field surveys like the LSST \citep{2012PASP..124.1175B,2018ApJS..236....9N}.



\acknowledgments
\section*{Acknowledgments}
YT and JW acknowledge the support of DFG priority program SPP 1992 ``Exploring the Diversity of Extrasolar Planets'' (WA 1047/11-1). RAS and EB gratefully acknowledge support from NASA grant NNX15AC97G. DMB acknowledges the support of the NYU Abu Dhabi Research Enhancement Fund under grant RE124. KH acknowledges support from STFC grant ST/R000824/1. This work was partly supported by the National Science Foundation of China (Grant No. 11333003, 11390372 and 11761131004 to SM). This work makes use of observations from the LCOGT network. This research uses data obtained through the Telescope Access Program (TAP), which has been funded by the National Astronomical Observatories of China, the Chinese Academy of Sciences, and the Special Fund for Astronomy from the Ministry of Finance. This work was partly supported by the National Science Foundation of China (Grant No. 11390372 and 11761131004 to SM).

\bibliography{tsapras}%






\end{document}